\pdfoutput=1

\documentclass[11pt,a4paper]{article}

\usepackage{jcappub}
\usepackage[utf8]{inputenc}
\usepackage{amsmath}
\usepackage{amsthm}
\usepackage{amssymb}
\usepackage{enumitem}   
\usepackage[english]{babel}
\usepackage{url}
\usepackage{mathtools}
\usepackage{slashed}
\usepackage{multirow}
\usepackage{lipsum}
\usepackage{xcolor}

\begin{document}


\hfill CERN-TH-2020-079


\title{LISA Sensitivity to Gravitational Waves from Sound Waves}
\author{Kai~Schmitz}
\affiliation{Theoretical Physics Department, CERN, 1211 Geneva 23, Switzerland}
\emailAdd{kai.schmitz@cern.ch}


\abstract{\textit{Gravitational waves} (GWs) produced by sound waves in the primordial plasma during a strong first-order phase transition in the early Universe are going to be a main target of the upcoming \textit{Laser Interferometer Space Antenna} (LISA) experiment.
In this short note, I draw a global picture of LISA's expected sensitivity to this type of GW signal, based on the concept of \textit{peak-integrated sensitivity curves} (PISCs) recently introduced in Refs.~\cite{Alanne:2019bsm,Schmitz:2020syl}.
In particular, I use LISA's PISC to perform a systematic comparison of several thousands of benchmark points in ten different particle physics models in a compact fashion.
The presented analysis (i) retains the complete information on the optimal signal-to-noise ratio, (ii) allows for different power-law indices describing the spectral shape of the signal, (iii) accounts for galactic confusion noise from compact binaries, and (iv) exhibits the dependence of the expected sensitivity on the collected amount of data.
An important outcome of this analysis is that, for the considered set of models, galactic confusion noise typically reduces the number of observable scenarios by roughly a factor two, more or less independent of the observing time.
The numerical results presented in this paper are also available on Zenodo~\cite{Schmitz:2020aaa}.}


\maketitle


\section{Introduction}
\label{sec:introduction}


Being the first experiment that is going to measure \textit{gravitational waves} (GWs) in the milli-Hertz range, the \textit{Laser Interferometer Space Antenna} (LISA)~\cite{Audley:2017drz,Baker:2019nia} is set to open a new observational window onto the Universe.
Among the possible GW signals in the LISA frequency band, the stochastic GW background~\cite{Maggiore:1999vm,Romano:2016dpx,Caprini:2018mtu,Christensen:2018iqi} from a \textit{strong first-order phase transition} (SFOPT) in the early Universe~\cite{Weir:2017wfa,Mazumdar:2018dfl} represents a particularly well-motivated example that is expected in many extensions of the standard model of particle physics (see, \textit{e.g.}, the two review reports by the LISA Cosmology Working Group~\cite{Caprini:2015zlo,Caprini:2019egz}).
In this paper, we will discuss LISA's projected sensitivity to this type of signal from a bird's eye view, focusing on the contribution from plasma sound waves that are generated during the phase transition~\cite{Hindmarsh:2013xza,Giblin:2013kea,Giblin:2014qia,Hindmarsh:2015qta,Hindmarsh:2016lnk,Hindmarsh:2017gnf,Hindmarsh:2019phv}.
Other possible contributions to the GW signal from a cosmological phase transition include the collision of vacuum bubbles~\cite{Kosowsky:1991ua,Kosowsky:1992rz,Kosowsky:1992vn,Kamionkowski:1993fg,Caprini:2007xq,Huber:2008hg} and magnetohydrodynamic turbulence~\cite{Caprini:2006jb,Kahniashvili:2008pf,Kahniashvili:2008pe,Kahniashvili:2009mf,Caprini:2009yp,Kisslinger:2015hua}.
The former, however, is negligibly small in most cases, while the latter presently still demands a better theoretical understanding.
Restricting ourselves to the sound-wave contribution to the GW signal therefore represents a conservative approach (see also the discussion in Ref.~\cite{Caprini:2019egz}). 


The main goal of this paper is to draw a global picture of LISA's sensitivity to the acoustic GW signal from a SFOPT across a large range of different particle physics models.
Remarkably enough, the standard approaches to this problem suffer from various limitations that make it difficult to scale up the number of benchmark points and\,/\,or models that are to be included in the analysis.
A simple comparison of theoretically computed GW spectra with LISA's power-law-integrated sensitivity curve~\cite{Thrane:2013oya}, \textit{e.g.}, does not preserve any information on the expected \textit{signal-to-noise ratio} (SNR), simply because the signal is not a pure power law, and quickly becomes impractical for a large number of spectra.
An analysis of the expected SNR as a function of the underlying SFOPT parameters $\alpha$, $\beta/H$, $T_*$, etc. (see Sec.~\ref{sec:signal}), on the other hand, is complicated by the high dimensionality $d$ of parameter space. 
In practice, one therefore often resorts to hypersurfaces in parameter space, keeping $d-2$ parameters fixed at characteristic values that roughly approximate the true values for the individual benchmark points that one is interested in.
This procedure may be useful as long as one considers one model at a time; however, it becomes impractical as soon as one intends to compare different models to each other that live on vastly separated hypersurfaces in parameter space.
In this paper, in order to avoid these limitations, we will therefore make use of the concept of \textit{peak-integrated sensitivity curves} (PISCs) that was recently introduced in Refs.~\cite{Alanne:2019bsm,Schmitz:2020syl}.
The key idea behind the PISC approach is to project the expected sensitivity of a given experiment not into the space of SFOPT parameters but into the space of observables that will eventually be part of the experimental data analysis. 
The acoustic GW signal from a SFOPT is primarily described by two such observables: (i) a peak frequency $f_{\rm peak}$, where the GW amplitude reaches a local maximum, and (ii) the GW amplitude at this frequency itself, $\Omega_{\rm peak}$, or equivalently, the integrated GW energy density, $\Omega_{\rm tot}$ (see Sec.~\ref{sec:signal}).
The PISC that will discuss in this paper, LISA's PISC for acoustic GWs, is thus defined as the experimental sensitivity curve in the two-dimensional parameter space spanned by $f_{\rm peak}$ and $\Omega_{\rm tot}$.
As we will show, this curve retains the complete information on the expected SNR and can be constructed in a universal fashion without any reference to a particular model or a specific hypersurface in parameter space.
Thanks to these characteristic properties, it serves as an ideal starting point for a systematic comparison of a large number of SFOPT scenarios.


For concreteness, we will consider in this paper the ten particle physics models that were discussed in Ref.~\cite{Caprini:2019egz} (see Sec.~\ref{sec:sensitivity}).
In particular, we will reproduce the complete information on the expected SNR for all 3720 individual benchmark points that were studied in Ref.~\cite{Caprini:2019egz} in a single PISC plot.
We will then demonstrate how this PISC plot can be used to address several questions for all ten models at the same time.
Specifically, we will discuss (i) the importance of \textit{galactic confusion noise} (GCN) from compact binaries~\cite{Nelemans:2004qz,Toonen:2012jj,Korol:2017qcx,Cornish:2017vip,Littenberg:2020bxy}, (ii) the time dependence of LISA's sensitivity as the mission progresses, and (iii) the impact of varying the spectral shape of the signal.
We argue that it is easier and more intuitive to address these questions based on our PISC plot rather than ten model-dependent SNR contour plots on ten different hypersurfaces in parameter space.
An important result of our analysis is that roughly half of all SFOPT scenarios that one would claim to be within LISA's reach in the absence of any GCN actually fail to pass the necessary SNR threshold when GCN is taken into account.
We thus conclude that, although this is presently not yet standard practice, GCN should always be accounted for in phenomenological studies of GWs from a SFOPT.


The rest of this paper is organized as follows.
In Sec.~\ref{sec:signal}, we will review the computation of the acoustic GW signal from a SFOPT, commenting in particular on the issue of shock formation in the plasma;
in Sec.~\ref{sec:noise}, we will compute LISA's strain noise power spectrum, making use of a novel analytical expression for LISA's signal response function;
and in Sec.~\ref{sec:sensitivity}, we will finally combine all ingredients, construct LISA's PISC, present our PISC master plot, and address the three questions listed above.
Sec.~\ref{sec:conclusions} contains our conclusions.


\section{Signal}
\label{sec:signal}


The spectrum of acoustic GWs produced during a SFOPT can be written as
\begin{equation}
\label{eq:signal}
\Omega_{\rm signal}\left(f\right) = \Omega_{\rm tot}\left(\alpha,\beta/H_*,T_*,v_w\right)\mathcal{S}\left(f/f_{\rm peak}\right) \,.
\end{equation}
In the following, we will now briefly introduce the different quantities in this expression; a more detailed discussion can be found in Refs.~\cite{Schmitz:2020syl,Caprini:2015zlo,Caprini:2019egz}.
The frequency-independent factor on the right-hand side of Eq.~\eqref{eq:signal}, $\Omega_{\rm tot}$, denotes the total energy density of acoustic GWs produced during the phase transition, in units of the critical energy density, $\rho_{\rm crit} = 3\,M_{\rm Pl}^2H^2$, 
\begin{equation}
\label{eq:Otot}
\Omega_{\rm tot} = \min\left\{1,H_*\tau_{\rm sh}\right\} \times 3\,\left(\frac{g_\rho^*}{g_\rho^0}\right)\left(\frac{g_s^0}{g_s^*}\right)^{4/3} \Omega_\gamma^0\,\tilde{\Omega} \left(8\pi\right)^{1/3}\frac{\max\left\{c_s,v_w\right\}}{\beta/H_*}\,K^2 \,.
\end{equation}
Here, $g_\rho$ and $g_s$ count the effective numbers of relativistic degrees of freedom that contribute to the radiation energy density $\rho_{\rm rad}$ and entropy density $s_{\rm rad}$, respectively.
The superscript on these quantities indicates whether they are evaluated at the present temperature $T_0$ or at the percolation temperature $T_*$, which represents the relevant temperature scale for GW production during the phase transition.
$\Omega_\gamma^0 \simeq 2.47\times10^{-5}/h^2$ quantifies the current photon energy density, where $h$ is the dimensionless Hubble parameter in the present epoch, which is defined via the relation $H_0 = 100\,h\,\textrm{km}/\textrm{s}/\textrm{Mpc}$. 
$\tilde{\Omega}$ quantifies the efficiency of GW production from sound waves and follows from integrating the shear stress unequal-time correlator of the bulk fluid~\cite{Hindmarsh:2015qta}.
The numerical simulations in Ref.~\cite{Hindmarsh:2017gnf} show that $\tilde{\Omega}$ is approximately constant for weak phase transitions.
Below, we will use $\tilde{\Omega} = 0.012$ as a representative value, which corresponds to one of the benchmark scenarios studied in  Ref.~\cite{Hindmarsh:2017gnf}.
The combination $\left(8\pi\right)^{1/3}\max\left\{c_s,v_w\right\}/\beta \equiv R_*$ on the right-hand side of Eq.~\eqref{eq:Otot} represents the mean bubble size, or equivalently, mean bubble separation at the time of percolation.%
\footnote{Describing the colliding bubbles by a regular cubic lattice of nonoverlapping spheres, the bubble number density at $T=T_*$ is given by $n_B = 1/R_*^3$.
Other lattice packings result in slight modifications of this relation.}
Here, $c_s$ denotes the speed of sound, $v_w$ is the bubble wall velocity in the plasma rest frame, and $\beta$ is a measure for the (inverse of the) duration of the phase transition.
The factor $K$ describes the energy fraction that is converted to kinetic energy of the bulk fluid during the phase transition,
\begin{equation}
\label{eq:K}
K = \frac{\kappa\,\alpha}{1+\alpha} \,,
\end{equation}
where $\alpha$ characterizes the strength of the phase transition and $\kappa$ is an efficiency factor that can be computed as a function of $\alpha$ and $v_w$.
In our analysis, we will follow Refs.~\cite{Caprini:2015zlo,Caprini:2019egz} and use the semianalytical fit functions for $\kappa$ that are provided in Appendix A of Ref.~\cite{Espinosa:2010hh}.


In passing, we mention that the derivation of Eq.~\eqref{eq:K} in Ref.~\cite{Espinosa:2010hh} is based on the so-called bag equation of state, which assumes a relativistic plasma with speed of sound $c_s^2 = 1/3$ in both the broken and the symmetric phase.
This assumption, however, may break down in realistic scenarios; see Ref.~\cite{Giese:2020rtr} for a recent analysis that re-evaluates the efficiency factor $\kappa$ for detonations with $c_s^2 \neq 1/3$ in the broken phase.
At present, the exact speed-of-sound profile across the bubble wall is unfortunately not known for most models.
In the following, we will therefore stick to the results of Ref.~\cite{Espinosa:2010hh} and set $c_s^2 = 1/3$ throughout our analysis.


The overall prefactor $ \min\left\{1,H_*\tau_{\rm sh}\right\}$ in Eq.~\eqref{eq:Otot} accounts for the finite lifetime of the GW source.
As stressed in Refs.~\cite{Ellis:2018mja,Ellis:2019oqb,Ellis:2020awk}, the generation of GWs from sound waves shuts off after a period $\tau_{\rm sh}$, when shocks begin to form, causing the motion of the bulk plasma to turn turbulent.
This leads to a suppression of the GW signal from sound waves whenever the time scale $\tau_{\rm sh}$ is shorter than a Hubble time.
$\tau_{\rm sh}$ can be estimated in terms of the mean bubble separation $R_*$ and the enthalpy-weighted root-mean-square of the fluid velocity, $\bar{U}_f$,
\begin{equation}
\tau_{\rm sh} = \frac{R_*}{\bar{U}_f} = \left(8\pi\right)^{1/3}\,\frac{\max\left\{c_s,v_w\right\}}{\beta\,\bar{U}_f} \,.
\end{equation}
$\bar{U}_f$ is given in terms of the kinetic-energy fraction $K$ and the mean adiabatic index $\Gamma$, which is defined as the ratio of the mean enthalpy density $\bar{w}$ and the mean energy density $\bar{e}$, 
\begin{equation}
\bar{U}_f = \left(\frac{K}{\Gamma}\right)^{1/2} = \left(\frac{3}{4}\,\frac{\kappa\,\alpha}{1+\alpha}\right)^{1/2} \,,
\end{equation}
where we used that $\Gamma = \bar{w}/\bar{e} = 4/3$ for a relativistic fluid.
Alternatively, one may estimate $\tau_{\rm sh}$ directly in terms of $R_*$ and $K$ and neglect the mean adiabatic index and the detour via the mean fluid velocity, $\tau_{\rm sh} = R_*/K^{1/2}$.
This is the approach adopted in Ref.~\cite{Caprini:2019egz}.
Another difference between Ref.~\cite{Caprini:2019egz} and our analysis is that we include a factor $3$ in our expression for $\Omega_{\rm tot}$, which is consistent with Ref.~\cite{Hindmarsh:2017gnf} (see, in particular, the erratum to this paper), whereas Ref.~\cite{Caprini:2019egz} includes a factor $1/c_s$, which renders the GW signal smaller by a factor $\sqrt{3}$.


The function $\mathcal{S}$ in Eq.~\eqref{eq:signal} describes the spectral shape of the GW signal.
One typically expects the spectrum to be peaked at a characteristic frequency $f_{\rm peak}$, which is why it is most convenient to write $\mathcal{S}$ as a function of $f/f_{\rm peak}$.
The peak frequency $f_{\rm peak}$ itself is set by the mean bubble separation $R_*$.
Its appropriately redshifted value in the present Universe reads 
\begin{equation}
f_{\rm peak} \simeq 8.9\times10^{-3}\,\textrm{mHz}\:\bigg(\frac{z_{\rm peak}}{10}\bigg)\left(\frac{\beta/H_*}{\max\left\{c_s,v_w\right\}}\right)\left(\frac{100}{g_s^*}\right)^{1/3}\left(\frac{g_\rho^*}{100}\right)^{1/2}\left(\frac{T_*}{100\,\textrm{GeV}}\right) \,.
\end{equation}
Here, $z_{\rm peak}$ characterizes the hierarchy between $R_*$ and the peak frequency at the time of the phase transition.
Its value needs to be inferred from numerical simulations, which indicate that it is generally close to $z_{\rm peak} = 10$~\cite{Hindmarsh:2017gnf}.
The spectral shape function $\mathcal{S}$ is often approximated by a broken power law that scales like $f^3$ at low frequencies and $f^{-4}$ at high frequencies.
This is, in particular, the ansatz chosen by the LISA Cosmology Working Group in Refs.~\cite{Caprini:2015zlo,Caprini:2019egz}.
However, a broken power law of this form is not always necessarily the best choice.
The analytical sound shell model of GW production from sound waves in Refs.~\cite{Hindmarsh:2016lnk,Hindmarsh:2019phv}, \textit{e.g.}, predicts an $f^5$ power law at low frequencies and an $f^{-3}$ power law at high frequencies.
Similarly, the outcome of numerical simulations is sometimes better described by an $f^{-3}$ fit rather than an $f^{-4}$ fit at high frequencies~\cite{Hindmarsh:2017gnf}.
Therefore, in order to account for this variability and uncertainty in the spectral shape function $\mathcal{S}$, we will go beyond the simple $f^3\rightarrow f^{-4}$ ansatz and consider a more general class of broken power laws in this paper,
\begin{equation}
\label{eq:S}
\mathcal{S}\left(x\right) = \frac{1}{\mathcal{N}}\:\mathcal{\widetilde{S}}\left(x\right) \,,\quad \mathcal{\widetilde{S}}\left(x\right) = \frac{x^p}{\left[q/\left(p+q\right) + p/\left(p+q\right) x^n\right]^{\left(p+q\right)/n}} \,.
\end{equation}
This function is constructed such that it exhibits the following characteristic properties,
\begin{equation}
\mathcal{S}'\left(x=1\right) = 0 \,, \quad \mathcal{S}\left(x=1\right) = \frac{1}{\mathcal{N}} \,, \quad \mathcal{S}\left(x\ll1\right) \propto f^p \,, \quad \mathcal{S}\left(x\gg1\right) \propto f^{-q} \,.
\end{equation}
We choose the normalization constant $\mathcal{N}$ such that the integral over $\mathcal{S}$ is normalized to unity,
\begin{align}
\label{eq:N}
\mathcal{N} = \int_{-\infty}^{+\infty} d\left(\ln x\right)\:\mathcal{\widetilde{S}}\left(x\right) = \left(\frac{q}{p}\right)^{p/n} \left(\frac{p+q}{q}\right)^{\left(p+q\right)/n} \frac{\Gamma\left(p/n\right)\Gamma\left(q/n\right)}{n\,\Gamma\left(\left(p+q\right)/n\right)} \,.
\end{align}
These definitions also allow us to rewrite Eq.~\eqref{eq:signal} in terms of the peak amplitude $\Omega_{\rm peak}$,
\begin{equation}
\Omega_{\rm signal}\left(f\right) = \Omega_{\rm peak}\left(\alpha,\beta/H_*,T_*,v_w,p,q,n\right)\mathcal{\widetilde{S}}\left(f/f_{\rm peak}\right) \,, \quad \Omega_{\rm peak} = \frac{1}{\mathcal{N}}\:\Omega_{\rm tot} \,.
\end{equation}
which illustrates that, unlike the total GW energy density parameter $\Omega_{\rm tot}$, the peak amplitude $\Omega_{\rm peak}$ also depends on the values of $p$, $q$, and $n$.
In our analysis, we will therefore keep working with Eq.~\eqref{eq:signal} and $\Omega_{\rm tot}$, such that the entire dependence on $p$, $q$, and $n$ is contained in $\mathcal{S}$.


In our notation, the standard ansatz for the spectral shape function in Refs.~\cite{Caprini:2015zlo,Caprini:2019egz} corresponds to $\left(p,q,n\right) = \left(3,4,2\right)$.
In this case, the normalization constant $\mathcal{N}$ is given by
\begin{equation}
\mathcal{N}\left(p=3,q=4,n=2\right) = \frac{343\,\sqrt{7}}{360\,\sqrt{3}} \simeq 1.455 \simeq \frac{1}{0.687} \,.
\end{equation}
In this paper, we will by contrast consider a larger set of $\left(p,q,n\right)$ tuples.
More specifically, we will work with $p\in\left\{3,4,5\right\}$, $q\in\left\{3,4\right\}$ and $n \in\left\{1,2,3\right\}$.
Here, our choice of $p$ and $q$ values reflects the range of power laws discussed further above, while the three discrete $n$ values allow us to describe a broad, a mid-sized, and a narrow peak in the spectrum, respectively.
In this way, we are able to cover a large range of possible spectral shapes and hence implicitly account for the dependence of $p$, $q$, and $n$ on the underlying properties of the phase transition.


\section{Noise}
\label{sec:noise}


Next, we turn to LISA's noise spectrum $\Omega_{\rm noise}$.
Again we will restrict ourselves to a brief summary of the basic ingredients; a more comprehensive review can be found in Appendix A of Ref.~\cite{Schmitz:2020syl}.
$\Omega_{\rm noise}$ is defined in terms of LISA's single-sided strain noise spectrum $S_{\rm noise}$,
\begin{equation}
\Omega_{\rm noise}\left(f\right) = \frac{2\pi^2}{3H_0^2}\,f^3 S_{\rm noise}\left(f\right) \,.
\end{equation}
In our analysis in this paper, we will consider two independent contributions to $S_{\rm noise}$: LISA's intrinsic instrumental strain noise $S_{\rm inst}$ as well as confusion noise from galactic binaries, $S_{\rm gcn}$, 
\begin{equation}
S_{\rm noise}\left(f\right) = S_{\rm inst}\left(f\right) + S_{\rm gcn}\left(f\right) 
\end{equation}
$S_{\rm inst}$ is in turn given by the detector noise spectrum $D_{\rm inst}$ and signal response function $\mathcal{R}$,
\begin{equation}
\label{eq:Sinst}
S_{\rm inst}\left(f\right) = \frac{D_{\rm inst}\left(f\right)}{\mathcal{R}\left(f\right)} \,.
\end{equation}


In the following, we shall now discuss the quantities $D_{\rm inst}$, $\mathcal{R}$, and $S_{\rm gcn}$ one after another.
The detector noise spectrum can be written as a sum of two stationary contributions~\cite{Cornish:2018dyw},%
\footnote{For a recent discussion of nonstationary noise sources for the LISA mission, see Ref.~\cite{Edwards:2020tlp}.}
\begin{equation}
\label{eq:DLISA}
D_{\rm inst}\left(f\right) = \frac{1}{L^2}\,D_{\rm oms}\left(f\right) + \frac{2}{\left(2\pi f\right)^4L^2}\left[1 + \cos^2\left(\frac{f}{f_*}\right)\right] D_{\rm acc}\left(f\right) \,.
\end{equation}
Here, $L = 2.5\times10^9\,\textrm{m}$ and $f_* = c/\left(2\pi L\right) \simeq 19.09\,\textrm{mHz}$ denote LISA's arm length and transfer frequency, respectively; $D_{\rm oms}$ and $D_{\rm acc}$ account for the noise in the \textit{optical metrology system} (OMS) (\textit{i.e.}, position noise) and the acceleration noise of a single test mass, respectively,
\begin{align}
D_{\rm oms}\left(f\right) & \simeq \left(1.5 \times 10^{-11}\,\textrm{m}\right)^2 \left[1 + \left(\frac{2\,\textrm{mHz}}{f}\right)^4\right] \textrm{Hz}^{-1} \,, \\\nonumber
D_{\rm acc}\left(f\right) & \simeq \left(3\times10^{-15}\,\textrm{m}\,\textrm{s}^{-2}\right)^2 \left[1 + \left(\frac{0.4\,\textrm{mHz}}{f}\right)^2\right] \left[1 + \left(\frac{f}{8\,\textrm{mHz}}\right)^4\right] \textrm{Hz}^{-1} \,.
\end{align}


LISA's signal response function $\mathcal{R}$ describes the response of an equal-arm Michelson interferometer to an incoming GW tensor mode.
In the past, one had to rely on semianalytical expressions or numerical techniques to evaluate this function~\cite{Larson:1999we,Larson:2002xr,Tinto:2010hz,Blaut:2012zz,Liang:2019pry,Zhang:2019oet}; only Ref.~\cite{Lu:2019log} recently succeeded in deriving a closed analytical expression for $\mathcal{R}$.
In our analysis, we will use this new analytical result; in particular, we will work with the compact form presented in Ref.~\cite{Zhang:2020khm},
\begin{align}
\label{eq:R}
u^2\,\mathcal{R}\left(u,\gamma\right)
& = s_{2u}\left[s_{\gamma/2}^2\left(\frac{1}{u}+\frac{2}{u^3}\right) + c_{\gamma/2}^2\left(2\,\textrm{Si}\left(2u\right) - \textrm{Si}\left(2u_+\right) - \textrm{Si}\left(2u_-\right) \vphantom{\ln c_{\gamma/2}^2} \right)\right]
\\ \nonumber
& \, + c_{2u}\left[s_{\gamma/2}^2\left(\frac{1}{6}-\frac{2}{u^2}\right) + c_{\gamma/2}^2\left(2\,\textrm{Ci}\left(2u\right) - \textrm{Ci}\left(2u_+\right) - \textrm{Ci}\left(2u_-\right) + \ln c_{\gamma/2}^2\right)\right]
\\ \nonumber
& \, - \frac{s_{u_+-u_-}}{32u\,s_{\gamma/2}^3}\left(21-28\,c_\gamma+7\,c_{2\gamma} + \frac{3-c_\gamma}{u^2}\right) + \frac{c_{u_+-u_-}}{8u^2 s_{\gamma/2}^2}\left(1+s_{\gamma/2}^2 \vphantom{\frac{3}{u^2}} \right)
\\ \nonumber
& \, - 2\,s_{\gamma/2}^2\left(\textrm{Ci}\left(2u\right) - \textrm{Ci}\left(u_+-u_-\right) + \ln s_{\gamma/2}\right) + \frac{3-c_\gamma}{12} - \frac{1-c_\gamma}{u^2} \,,
\end{align}
where we introduced the following notation to arrive at an even more compact expression,
\begin{align}
s_x = \sin\left(x\right) \,, \quad c_x = \cos\left(x\right) \,, \quad u_\pm = u \pm u\,\sin\left(\frac{\gamma}{2}\right) \,.
\end{align}
The independent variables $u$ and $\gamma$ in Eq.~\eqref{eq:R} represent the GW frequency in units of the transfer frequency, $u = f/f_*$, and the opening angle of the interferometer, respectively.
In the case of LISA, which will consist of three spacecraft in an equilateral triangular formation, we have $\gamma = \pi/3$.
Finally, $\textrm{Si}$ and $\textrm{Ci}$ in Eq.~\eqref{eq:R} represent sine and cosine integral functions,
\begin{equation}
\textrm{Si}\left(x\right) = \int_0^x dt\:\frac{\sin\left(t\right)}{t} \,, \quad \textrm{Ci}\left(x\right) = -\int_x^\infty dt\:\frac{\cos\left(t\right)}{t} \,.
\end{equation}
By making use of Eq.~\eqref{eq:R} instead of, say, an approximate fit function for $\mathcal{R}$ (see, \textit{e.g.}, Ref.~\cite{Cornish:2018dyw}), we are able to eliminate all numerical uncertainties in computing $S_{\rm inst}$.
The net effect of this will be small.
On the other hand, there is no reason to artificially reduce the precision of our analysis when an exact result for $\mathcal{R}$ is readily available in the literature.
For a discussion of the signal response in the case of unequal interferometer arm lengths, which will ultimately correspond to the physical situation during the LISA mission, see also~\cite{Wang:2020fwa,Vallisneri:2020otf}.


The last ingredient entering the noise spectrum is the GCN from compact binaries, $S_{\rm gcn}$.
A semianalytical fit function for $S_{\rm gcn}$ has been worked out in Ref.~\cite{Cornish:2017vip}, which estimates the GCN seen by LISA based on the compact-binary population model in Refs.~\cite{Nelemans:2004qz,Toonen:2012jj},%
\footnote{Here, we correct a typo that appears in Eq.~(3) of Ref.~\cite{Cornish:2017vip} as well as in Eq.~(14) of Ref.~\cite{Cornish:2018dyw}.
In both expressions for $S_{\rm gcn}$, the sign in front of the coefficient $\beta$ needs to be negative, \textit{i.e.}, $+\beta$ needs to be replaced by $-\beta$.
We thank Travis Robson and Neil J.\ Cornish for a helpful discussion on this point.}
\begin{equation}
\label{eq:gcn}
S_{\rm gcn}\left(f\right) = A\,\left(\frac{1\,\textrm{mHz}}{f}\right)^{7/3} \exp\left[-\left(f/f_{\rm ref}\right)^\alpha - \beta f\sin\left(\kappa f\right)\right]\left[1 + \tanh\left(\gamma\left(f_{\rm knee}-f\right)\right)\right] \,.
\end{equation}
This fit has been obtained making use of a variant of the \textsc{BayesLine} algorithm~\cite{Littenberg:2014oda} that applies Markov chain Monte Carlo techniques to sample the entire parameter space of the compact-binary population model.
The expression in Eq.~\eqref{eq:gcn} therefore marginalizes over the parameters of the noise model, which allows us to directly compare and add it to the sky- and polarization-averaged instrumental noise in Eq.~\eqref{eq:Sinst}.
The overall amplitude $A$ and reference frequency $f_{\rm ref}$ in Eq.~\eqref{eq:gcn} are fixed at constant values, $A =  9 \times 10^{-38}\,\textrm{Hz}^{-1}$ and $f_{\rm ref} = 1000\,\textrm{mHz}$.
Meanwhile, the parameters $\alpha$, $\beta$, $\kappa$, and $\gamma$ as well as the frequency $f_{\rm knee}$, which describes the position of a knee-like feature in the GCN spectrum, depend on LISA's observing time; see Tab.~\ref{tab:confusion}.
The reason for this is that, as the mission progresses, the increasingly larger amount of data will allow one to successively resolve more and more compact binaries, which can then be individually subtracted from the unresolved GCN background.
As a result, the GCN contribution to LISA's strain noise spectrum, $S_{\rm gcn}$, decreases as function of $t_{\rm data}$.
In Sec.~\ref{sec:sensitivity}, we will explicitly account for this time dependence of the noise spectrum and consider the four benchmark values that were studied in Ref.~\cite{Cornish:2017vip}, $t_{\rm data} = 0.5,\,1.0,\,2.0,\,4.0\,\textrm{yr}$.


\begin{table}
\begin{center} 
\renewcommand{\arraystretch}{1.22}
\caption{Parameters appearing in the fit function for the GCN spectrum $S_{\rm gcn}$ in Eq.~\eqref{eq:gcn} as functions of the collected amount of data $t_{\rm data}$~\cite{Cornish:2017vip}.
Note that Ref.~\cite{Cornish:2017vip} works in units where $1\,\textrm{Hz}$ is set to $1$.}
\label{tab:confusion}
\medskip\smallskip
\begin{tabular}{|c||ccccc|}
\hline
$t_{\rm data}$ $[\textrm{yr}]$       &
$\alpha$                             &
$\beta$        $[\textrm{mHz}^{-1}]$ &
$\kappa$       $[\textrm{mHz}^{-1}]$ &
$\gamma$       $[\textrm{mHz}^{-1}]$ &
$f_{\rm knee}$ $[\textrm{mHz}]$                                 \\ 
\hline\hline
0.5 & $0.133$ & $\phantom{-}0.243$ & $0.482$ & $0.917$ & $2.58$ \\
1.0 & $0.171$ & $\phantom{-}0.292$ & $1.020$ & $1.680$ & $2.15$ \\
2.0 & $0.165$ & $\phantom{-}0.299$ & $0.611$ & $1.340$ & $1.73$ \\
4.0 & $0.138$ & $          -0.221$ & $0.521$ & $1.680$ & $1.13$ \\
\hline
\end{tabular}
\end{center} 
\end{table}


\section{Sensitivity}
\label{sec:sensitivity}


The SNR $\varrho$ for given signal and noise spectra $\Omega_{\rm signal}$ and $\Omega_{\rm noise}$ can be computed as~\cite{Maggiore:1999vm,Allen:1996vm,Allen:1997ad},
\begin{equation}
\label{eq:snr}
\varrho = \left[t_{\rm data} \int_{f_{\rm min}}^{f_{\rm max}}df \left(\frac{\Omega_{\rm signal}\left(f\right)}{\Omega_{\rm noise}\left(f\right)}\right)^2\right]^{1/2} \,.
\end{equation}
This expression is valid for an idealized auto-correlation measurement with perfect noise subtraction, assuming a stochastic, Gaussian, stationary, isotropic, and unpolarized GW background in the weak-signal regime.
The SNR in Eq.~\eqref{eq:snr} moreover corresponds to the \textit{optimal} SNR that can in principle be achieved when an optimally chosen matched filter is applied to the data.
This is reflected in the fact that Eq.~\eqref{eq:snr} assumes knowledge of the signal spectrum $\Omega_{\rm signal}$ that one actually attempts to measure.
The upshot of these caveats is that Eq.~\eqref{eq:snr} should be regarded as the theoretical \textit{upper bound} on the expected SNR.
If LISA should indeed detect a GW signal from a cosmological phase transition, the actual SNR based on the real data is likely going to be smaller (see, \textit{e.g.}, the $\chi^2$ analysis in Ref.~\cite{Kuroyanagi:2018csn}).%
\footnote{Another important factor is the efficiency of compact-binary subtraction in the course of the LISA mission.
On the one hand, one might not be able to individually resolve and subtract as many compact binaries as initially expected.
On the other hand, additional characteristics of the GCN signal, such as its anisotropy and time dependence, might in fact help in its subtraction.
We leave a more detailed investigation for future work; in this paper, we are going to follow Refs.~\cite{Cornish:2017vip,Cornish:2018dyw} and work with the noise model described in Sec.~\ref{sec:noise}.}


The key idea behind the PISC concept is to rewrite the SNR in Eq.~\eqref{eq:snr} as follows~\cite{Alanne:2019bsm,Schmitz:2020syl},
\begin{equation}
\varrho = \frac{\Omega_{\rm tot}\left(\alpha,\beta/H_*,T_*,v_w\right)}{\Omega_{\rm pis}\left(f_{\rm peak},t_{\rm data},p,q,n\right)} \,,
\end{equation}
where the peak-integrated sensitivity $\Omega_{\rm pis}$ follows from integrating over the noise spectrum $\Omega_{\rm noise}$ after weighting it with the shape function $\mathcal{S}$ for the acoustic GW signal from a SFOPT,
\begin{equation}
\Omega_{\rm pis}\left(f_{\rm peak},t_{\rm data},p,q,n\right) = \left[t_{\rm data} \int_{f_{\rm min}}^{f_{\rm max}}df \left(\frac{\mathcal{S}\left(f/f_{\rm peak},p,q,n\right)}{\Omega_{\rm noise}\left(f,t_{\rm data}\right)}\right)^2\right]^{-1/2} \,.
\end{equation}
For a given experiment and fixed shape function, $\Omega_{\rm pis}$ can be constructed once and for all.
That is, in order to compute the SNR values for a set of benchmark points, it is no longer necessary to carry out the frequency integration in Eq.~\eqref{eq:snr} over and over again.
Instead, it suffices to construct the experimental PISC in the two-dimensional parameter space spanned by $f_{\rm peak}$ and $\Omega_{\rm tot}$ and focus on computing the theoretical predictions for exactly these two observables.
Each benchmark scenario can then be represented by its $\left(f_{\rm peak},\Omega_{\rm tot}\right)$ coordinates in this two-dimensional space, and the corresponding optimal SNR can be identified as the vertical distance $\Delta y$ between the point $\left(x,y\right) = \left(f_{\rm peak},\Omega_{\rm tot}\right)$ and the PISC of interest.
In other words, PISC plots retain the full information on the SNR and encode it on the $y$ axis.


\begin{figure}
\begin{center}
\includegraphics[width=\textwidth]{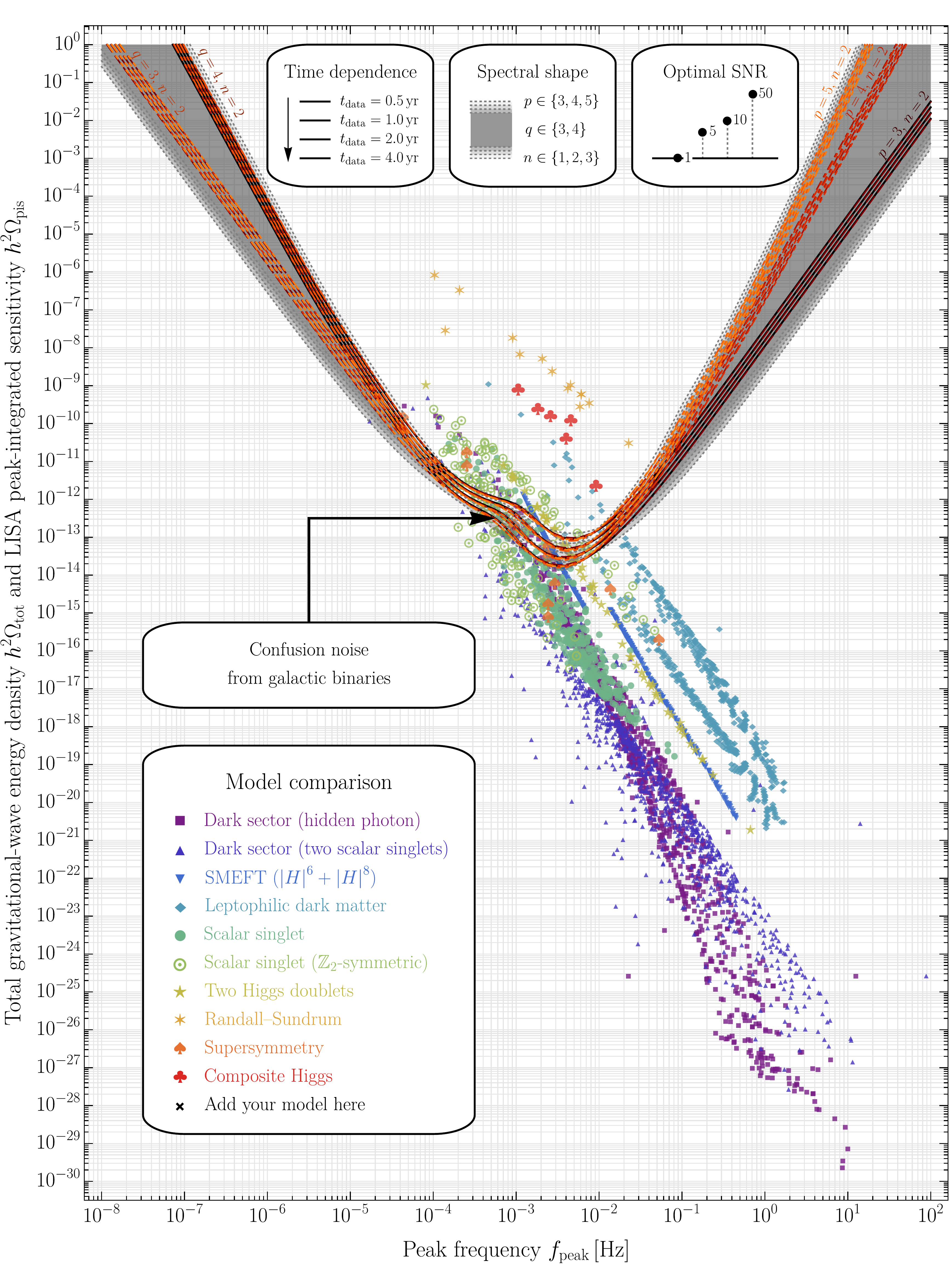}
\caption{Global picture of LISA's sensitivity to the acoustic GW signal from a SFOPT.
See text.}
\label{fig:1}
\end{center}
\end{figure}


\begin{figure}
\begin{center}
\includegraphics[width=\textwidth]{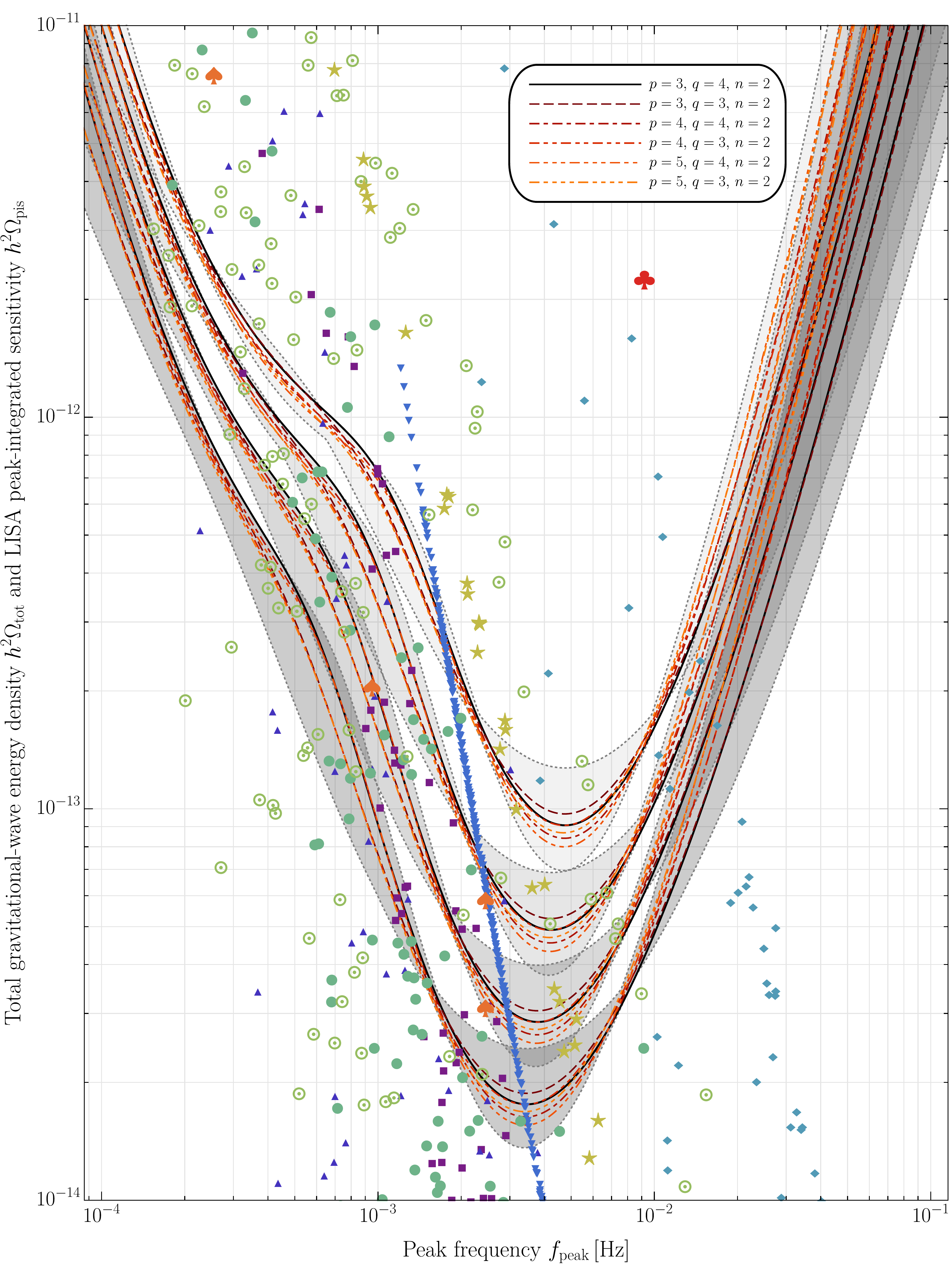}
\caption{Same as Fig.~\ref{fig:1}, zoomed into the most relevant region of the $f_{\rm peak}$\,--\,$\Omega_{\rm tot}$ parameter space.}
\label{fig:2}
\end{center}
\end{figure}


We are now ready to turn to the main part of our analysis and draw our PISC master plot for LISA's sensitivity to acoustic GWs from a cosmological phase transition; see Fig.~\ref{fig:1}.
In this plot, we show how LISA's PISC varies for different assumptions regarding the spectral shape of the signal\,---\,$p\in\left\{3,4,5\right\}$, $q\in\left\{3,4\right\}$, $n\in\left\{1,2,3\right\}$\,---\,and as a function of the observing time $t_{\rm data}$.
In addition, we indicate the $\left(f_{\rm peak},\Omega_{\rm tot}\right)$ coordinates of all 3720 benchmark points that were studied in Ref.~\cite{Caprini:2019egz}.
These benchmark points belong to the following ten models:
\begin{enumerate}
\item Dark-sector model featuring a spontaneously broken $U(1)$ gauge symmetry~\cite{Breitbach:2018ddu}
\item Dark-sector model featuring two gauge-singlet scalars~\cite{Breitbach:2018ddu}
\item $\left|H\right|^6$ and $\left|H\right|^8$ operators in the \textit{standard model effective field theory} (SMEFT)~\cite{Chala:2018ari}
\item Dark-matter model based on gauged and spontaneously broken lepton number~\cite{Madge:2018gfl}
\item Extension of the standard model by a real scalar singlet~\cite{Chen:2017qcz,Alanne:2019bsm}
\item Extension of the standard model by a $\mathbb{Z}_2$-symmetry-protected real scalar singlet~\cite{Huang:2016cjm,Beniwal:2017eik}
\item Two-Higgs-doublet model with a softly broken $\mathbb{Z}_2$ symmetry~\cite{Dorsch:2013wja,Dorsch:2014qja,Dorsch:2016nrg,Dorsch:2017nza}
\item Holographic phase transitions in extra-dimensional Randall--Sundrum models~\cite{Megias:2018sxv}
\item Selection of supersymmetric models featuring chiral singlet or triplet fields~\cite{Huber:2015znp,Garcia-Pepin:2016hvs,Bian:2017wfv,Demidov:2017lzf}
\item Composite-Higgs models featuring different pseudo-Nambu--Goldstone bosons~\cite{Bruggisser:2018mrt}
\end{enumerate}
More details on these models can be found in Ref.~\cite{Caprini:2019egz} and on the web page of the associated online tool \textsc{PTPlot}~\cite{ptplot:2019aaa}, which lists the SFOPT parameters for all 3720 benchmark points.%
\footnote{We thank the LISA Cosmology Working Group for explicitly giving us permission to use this data.}
A characteristic feature of Fig.~\ref{fig:1} (see also the enlarged section in Fig.~\ref{fig:2}) is that it allows us to present the expected SNR values of all benchmark points and models in a single plot.
We reiterate that this is infeasible when working instead with model-specific SNR plots on approximate hypersurfaces in the high-dimensional SFOPT parameter space (see Sec.~\ref{sec:introduction}).


Figs.~\ref{fig:1} and \ref{fig:2} provide a graphical illustration of how varying the observing time $t_{\rm data}$ and spectral shape of the signal affects LISA's sensitivity reach.
For each $\left(p,q,n\right)$ tuple, Figs.~\ref{fig:1} and \ref{fig:2} show four PISCs that indicate LISA's sensitivity reach after $t_{\rm data} = 0.5,\,1.0,\,2.0,\,4.0\,\textrm{yr}$ (from top to bottom), respectively.
The four times six curves for $n=2$ are drawn explicitly in both figures; the envelopes of all other curves are indicated by gray-shaded bands.
We moreover repeat that, for any given benchmark point, the distance between this point and any of the PISCs along the $y$ direction equals the corresponding optimal SNR.
This is also illustrated by the ruler in the top-right legend box in Figs.~\ref{fig:1}, which is true to scale.


In order to quantify the effect of varying the observing time or spectral shape of the signal, we shall now determine the number of benchmark scenarios that surpass an SNR threshold of $\varrho_{\rm thr} = 10$ in dependence of the parameters $p$, $q$, $n$, and $t_{\rm data}$.
Here, $\varrho_{\rm thr} = 10$ represents the detection threshold that was identified and used in the analysis of Ref.~\cite{Caprini:2015zlo}.
Of course, such a counting analysis is not going to replace a proper statistical analysis that would quantify the probability of LISA observing a signal in particular model or class of models.
However, given the present state of the art, which amounts to solitary benchmark points and independent collections of model scans, we believe that our simple counting analysis can at least convey a rough impression of the effect of different signal shapes and observing times. 

\begin{figure}
\begin{center}
\includegraphics[width=0.63\textwidth]{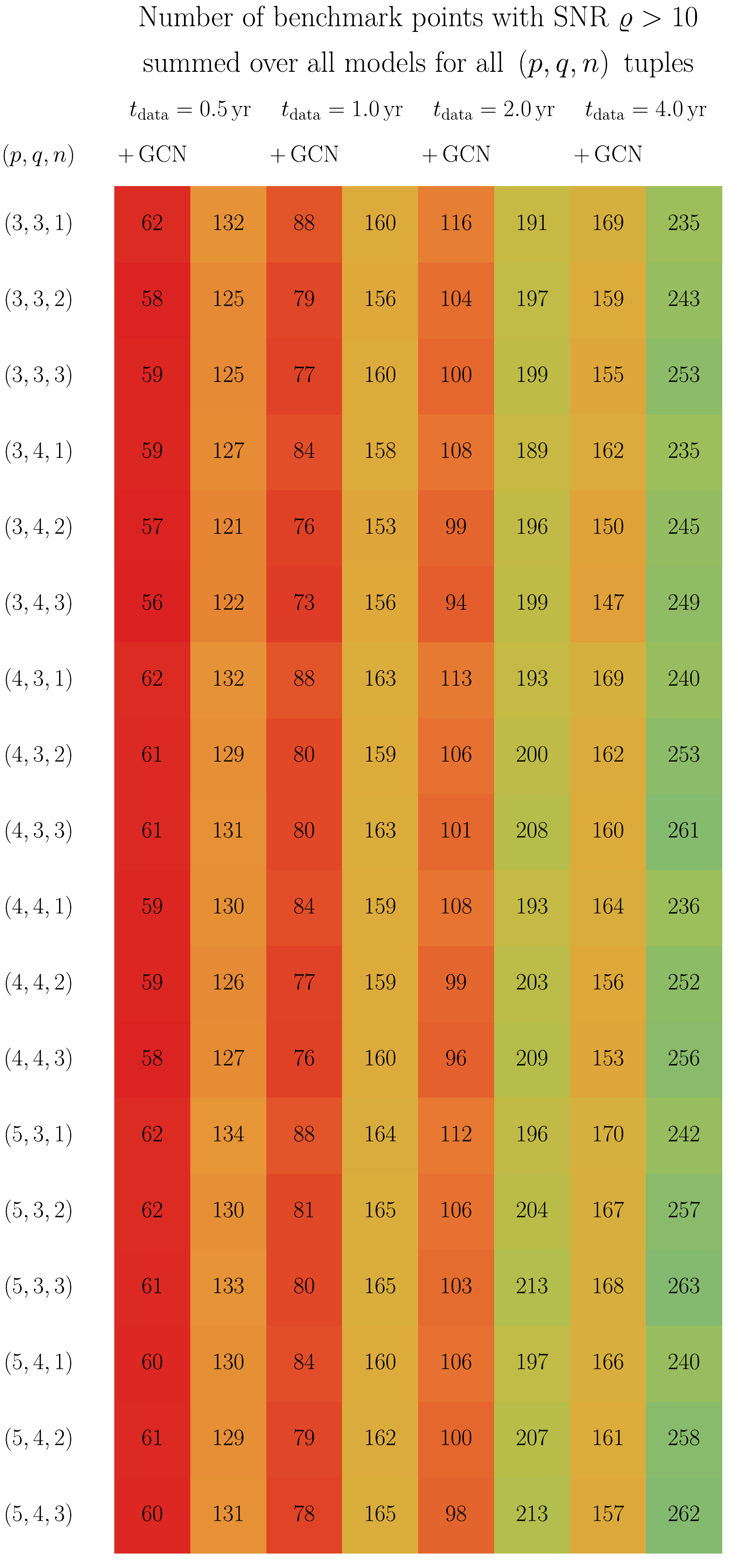}
\caption{Number of benchmark points above detection threshold in dependence of $p$, $q$, $n$, and $t_{\rm data}$.
The color code reflects the spectrum of values in this table, ranging from 56 (red) to 263 (green).}
\label{fig:3}
\end{center}
\end{figure}
\begin{figure}
\begin{center}
\includegraphics[width=0.895\textwidth]{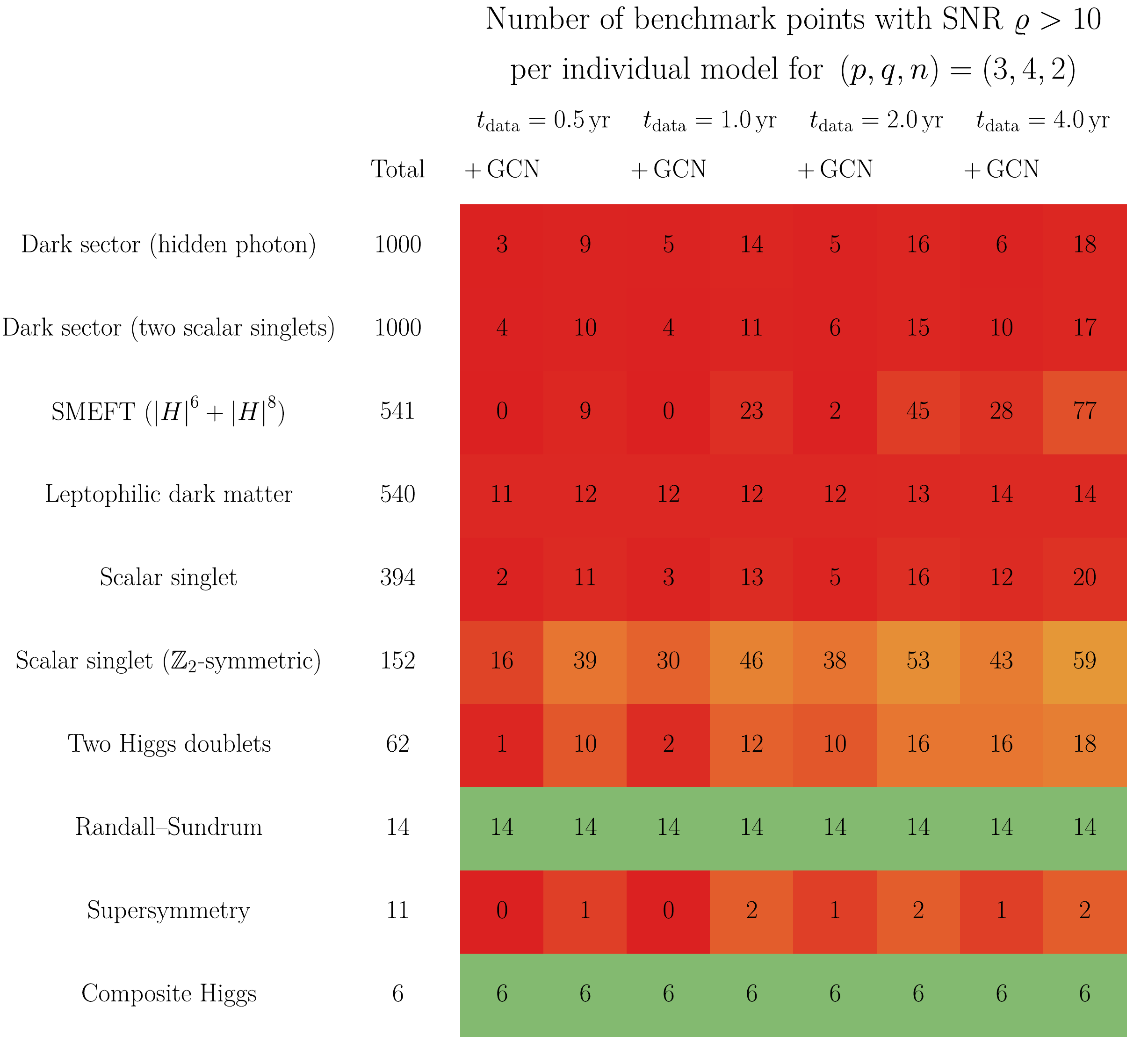}
\caption{Number of benchmark points above detection threshold for $\left(p,q,n\right) = \left(3,4,2\right)$.
The color code reflects the relative number of observable points, ranging from $0\,\%$ (red) to $100\,\%$ (green).}
\label{fig:4}
\end{center}
\end{figure}

In a first step, we count the total number of above-threshold scenarios without distinguishing between different models; see Fig.~\ref{fig:3}.
The general trend that can be read off from this figure is that, without accounting for GCN, an observing time of $t_{\rm data} = 0.5\,\textrm{yr}$ results in roughly 130 observable scenarios, which increases to roughly 250 observable scenarios after an observing time of $t_{\rm data} = 4.0\,\textrm{yr}$.
However, accounting for GCN in the analysis, these numbers are reduced by roughly a factor $2.2$ and $1.6$, respectively, down to roughly $60$ observable scenarios for $t_{\rm data} = 0.5\,\textrm{yr}$ and roughly 160 observable scenarios for $t_{\rm data} = 4.0\,\textrm{yr}$.
The spread in these numbers related to different choices of $p$, $q$, and $n$ is generally very small, which is also evident from the sensitivity curves shown in Figs.~\ref{fig:1} and \ref{fig:2}.
In a second step, we repeat our analysis for the standard spectral shape that is described by $\left(p,q,n\right) = \left(3,4,2\right)$; see Fig.~\ref{fig:4}.
Now we count the number of benchmark scenarios above detection threshold for each individual model. 
Again we find that including GCN has a significant impact on the number of observable benchmark points, although now we see that the effect can be very different from model to model.
Extremely strong phase transitions, such as those in Randall--Sundrum and composite-Higgs models, will simply be always observable.%
\footnote{At present, the description of these scenarios, however, relies on extrapolating the results of numerical simulations that were performed for weak phase transitions, $\alpha \lesssim 1$, across many orders of magnitude in $\alpha$.
The most promising benchmark points in the sample are therefore also the least understood.
More work is needed to improve our understanding of the dynamics and GW signature of these extreme phase transitions.}
The detectability of other models, on the other hand, significantly improves in the course of the LISA mission.
This is, \textit{e.g.}, the case for the $\mathbb{Z}_2$-symmetric real-scalar-singlet model, for which many benchmark points lie only slightly above the LISA PISC at the beginning of the mission.
These points notably benefit from the improving sensitivity as the collected amount of data increases.


\section{Conclusions}
\label{sec:conclusions}


In this short note, we presented a global picture of LISA's sensitivity to acoustic GWs from a SFOPT in the early Universe.
Our main results are the PISC plots in Figs.~\ref{fig:1} and \ref{fig:2}, which retain the complete information on the optimal SNR and visualize LISA's sensitivity to a broad range of SFOPT scenarios in a compact fashion.
We argued that these plots are well suited to collectively study a large number of benchmark points and models at the same time, which we demonstrated by discussing (i) the importance of GCN, (ii) the time dependence of LISA's PISC, and (iii) the impact of varying the spectral shape of the signal.

To illustrate the characteristic features of our approach, we also reproduced the predictions for all benchmark points and models that were discussed in the second review report by the LISA Cosmology Working Group~\cite{Caprini:2019egz}.
Out of the 3720 benchmark points in this dataset, roughly 250 points (\textit{i.e.}, roughly $7\,\%$) lead to an optimal SNR above detection threshold after an observing time of $t_{\rm data} = 4.0\,\textrm{yr}$ when GCN is neglected.
This number is reduced by roughly a factor of $2$ down to roughly 160 points (\textit{i.e.}, roughly $4\,\%$) when GCN is taken into account following the treatment in Refs.~\cite{Cornish:2017vip,Cornish:2018dyw} (see Fig.~\ref{fig:3}).
Of course, the situation may vary strongly from model to model (see Fig.~\ref{fig:4}).
In general, we are, however, able to to draw two basic conclusions:
(i) In most cases, only a small fraction of all benchmark points is above detection threshold; and those benchmark points that manage to surpass the detection threshold often times rely on extrapolating simulation results across many orders of magnitude.
(ii) The impact of GCN is nonnegligible and should always be accounted for in phenomenological studies of GWs from a cosmological phase transition.
In this work, we did this using the fit function derived in Refs.~\cite{Cornish:2017vip}, which marginalizes over the uncertainties of the GCN model.
In future work, it would be interesting to refine our analysis based on a more sophisticated treatment of GCN.
In summary, we conclude that the GW phenomenology of SFOPTs in the early Universe remains an exciting, but also challenging topic.
More work is needed to identify models with good detection prospects, \textit{i.e.}, models where the fraction of accessible benchmark points is above the percent level; and more work is needed to better simulate and understand the phase transitions in these models from a theoretical perspective.


\section*{Acknowledgments}


I wish to thank Sergei Bobrovskyi, Neil J.\ Cornish, John Ellis, Felix Giese, Max Hoffmann, Thomas Konstandin, Marek Lewicki, Toby Opferkuch, Travis Robson, Jorinde van de Vis, and David Weir for helpful discussions and comments.
I am also grateful to the chairs of the LISA Cosmology Working Group for giving me permission to use the data on \href{http://ptplot.org}{ptplot.org}.
This project has received funding from the European Union's Horizon 2020 Research and Innovation Programme under grant agreement number 796961, ``AxiBAU'' (K.\,S.).


\bibliographystyle{JHEP}
\bibliography{arxiv_3.bib}


\end{document}